\documentclass[twocolumn]{article}
\usepackage[latin1]{inputenc}
\usepackage{graphics}
\usepackage{graphicx}
\usepackage{dcolumn}
\usepackage{amsmath}
\usepackage{amssymb}

\newcommand{\bfs}{{\bf s}}
\newcommand{\bfp}{{\bf p}}
\newcommand{\bfq}{{\bf q}}
\newcommand{\bfr}{{\bf r}}

\newcommand{\bfsz}{{\bf s_{0}}}

\newcommand{\Dtt}{{\Delta t_{\rm true}}}
\newcommand{\Dtf}{{\Delta t_{\rm found}}}
\newcommand{\ts}{{t_{\rm s}}}

\begin{document}

\title{Fast analytical methods for the correction of signal random time-shifts and application to segmented HPGe detectors}

\author{P. Désesquelles$^1$, T.M.H. Ha$^1$, A. Korichi$^1$, F. Le Blanc$^2$,\\
 A. Olariu$^2$, C.M. Petrache$^2$\\
 \\
$^1$ CSNSM CNRS/IN2P3 and Université Paris Sud 11,\\
15 rue G. Clémenceau, 91405 Orsay, France\\
$^2$ IPNO CNRS/IN2P3 and Université Paris Sud 11,\\
15 rue G. Clémenceau, 91405 Orsay, France \\
On behalf of the AGATA Collaboration}

\maketitle

\begin{abstract}
Detection systems rely more and more on on-line or off-line comparison of detected signals with basis signals in order to determine the characteristics of the impinging particles. Unfortunately, these comparisons are very sensitive to the random time shifts that may alter the signal delivered by the detectors. We present two fast algebraic methods to determine the value of the time shift and to enhance the reliability of the comparison to the basis signals.
\end{abstract}

\section{Introduction}

Modern detection systems often rely on the on-line shape analysis of the signals delivered by the detectors. In this context, signal uncertainties due to noise, cross-talk or time jitter are of tremendous importance. This paper introduces new methods to address the latter effect.

In grid search algorithms \cite{Bev}, the detected signal is systematically compared to all the signals of a basis in order to find the best-match. The signals from the basis are considered as reference signals, they do not present any time shifts. The characteristics of the ``event'' that has generated the pulse are determined by searching the best-match among the basis signals. This technique is used in many modern applications, for example to determine the charge and mass of the detected ions \cite{Ham} or the numbers of carbon and hydrogen in atomic clusters \cite{Cha,Mar2}. We will illustrate the methods by the search of gamma-ray interaction locations \cite{Ola} in an HPGe crystal \cite{Sim,Lee}. A basis of signals corresponding to a set of regularly spaced hit locations covering the crystal volume is generated using a signal simulation code \cite{Med}, or a crystal scanning device \cite{Bos}. Locating an experimental gamma-ray interaction consists in finding the best-match between the delivered signal and the basis signals. This best-match is supposed to correspond to the point closest to the actual hit. When detected signals are altered by time shifts, this property may no longer be true.

Usual methods to correct random time shifts consist whether in determining a given point on the signal (e.g. the crossing of a threshold) or in comparing the experimental signal to versions of the reference signal translated by different $\Delta t$. In the former case, the comparison, relying on a single sample, is very sensitive to noise and would not allow determining time shifts smaller than the sample duration. In the latter case, the computation time is cripplingly long for most on-line applications.

We present in Section 2 a reliable and fast method to find the best-match to a signal affected by random time shifts among a basis of reference signals. This method does not give directly the value of the time shift. In Section 3, we present an algebraic fast method to both determine the time shift of the delivered signal and compare it to a signal basis.

\section{Comparison of  time-shifted signals with reference signals}

\subsection{Method}

In the following, the signal not affected by a time shift will be noted $s_{0}(t)$ where $t$ is the time index (sample number). To reduce the size of the equations, the signals will also be denoted as vectors (bold lower cases). The signal actually delivered by the detector, $\bfs$, is different from $\bfsz$ due to random noise and time shift. The squared residue between the reference signal and the corrected signal can be written as:

\begin{equation}
\label{Eq R2 Dt}
R^2 = \sum_{t} \left(s_{0}(t) - s(t) \right)^2=(\bfsz-\bfs)^{2}\,.
\end{equation}

\begin{figure}[htbp]
\begin{center}
\includegraphics[width=7cm]{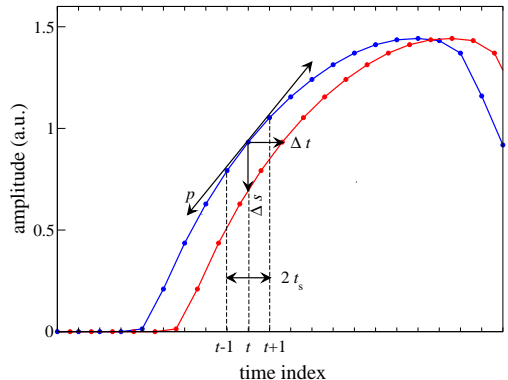}
\caption{The time shift $\Delta t$ induces an amplitude shift $\Delta s$ proportional to the slope $p$.}
\label{Fig Ds_pDt}
\end{center}
\end{figure}

The pulse shape analysis algorithms generally consist in comparing detected signals with a set of basis signals. The best-match criterion is usually the residue, but, when the data are affected by noise, fluctuations or uncertainties, the rigorous criterion is the chi-square:

\begin{equation}
\label{Eq chi2}
\chi^2 = \sum_{t} \frac{\left( s_{0}(t) - s(t)\right)^2}{\sigma^2_{\bfs}(t) }\ ,
\end{equation}

\noindent where  $\sigma_{\bfs}$ is the standard deviations of the uncertainties on $\bfs$. The choice of the residue criterion is justified by the fact that, most of the time, the signals are affected only by a white noise with a standard deviation identical for all samples. In this case the residue and the chi-square minimizations give the same results. However, when the uncertainties are not the same for all samples, the chi-square minimization must be chosen. This is the case when the signals are translated by a random, unknown, time shift. Where the signal is horizontal, if the shift is small enough, it has no influence on the signal amplitude. On the other hand, for rapid increases or rapid decreases, the biases on the sample values are high. In fact, the shift is proportional to the signal slope and to the time shift: $\Delta s(t) = -p(t)\,\Delta t$ as can be seen in Fig. \ref{Fig Ds_pDt}. The denominators of the chi-square terms are the quadratic sums of the uncertainties due to the noise and to the time jitter: 

\begin{equation}
\label{Eq sigma2s}
\sigma^2_{\bfs}(t) = \sigma^2_{\rm noise}+p^2(t)\,\sigma^2_{\Delta t}\,,
\end{equation}

\noindent where $\sigma_{\Delta t}$ is the standard deviation of the time shift, which value is supposed to be known. The estimate of the signal slope is given by:

\begin{equation}
\label{Eq p(t)}
p(t) =\frac{s_{0}(t+1)-s_{0}(t-1)}{2\,\ts}\,,
\end{equation}
\noindent where $\ts$ is the time width of the signal samples.

\subsection{Results}

\begin{figure}[htbp]
\begin{center}
\includegraphics[width=7cm]{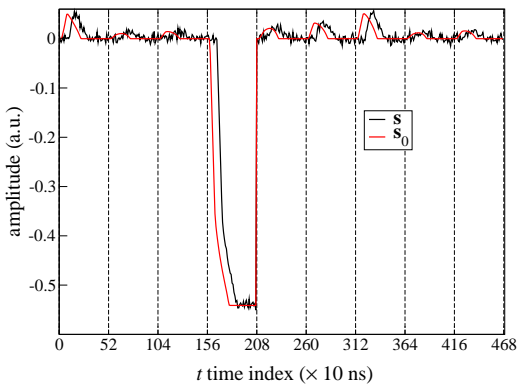}
\caption{Example of meta-signal obtained by the concatenating of the hit segment signal (fourth signal) and its eight neighbors signals. The red signal corresponds to the reference, the black line represent the test signal (the time shift is 75~ns and the noise is 1\% of the signal maximum).}
\label{Fig s_s0_340}
\end{center}
\end{figure}

The validation of the method will be performed using the AGATA segmented coaxial HPGe \cite{Sim} signal. A signal basis has been generated along a cubic grid of 2 mm step using the MGS code \cite{Med}. When a gamma-ray interacts with the crystal, signals are induced in the segment of the detector where the hit took place and in the neighboring segments. To handle the information as a whole, we introduce the so-called meta-signals, which are the concatenation of all the induced signals. An example is given in Fig. \ref{Fig s_s0_340}. The signal corresponding to the hit segment is the fourth one. The other eight signals are those induced in the direct neighbors of the hit segment. The time shift being the same for all segment signals, it also applies to meta-signals. In the following, the $j^{\rm th}$ basis meta-signal will be noted $\bfs_{j}$. The basis meta-signals play the role of the reference signals. The test meta-signals are also generated with the MGS code. They will be noted $\bfs$. The sampling rate for basis and test signals corresponds to $\ts=10$~ns samples. However, the test signals are first generated on 1~ns bins so that they can be moved by time shifts smaller than the sample time, then grouped into 10~ns samples.

\begin{figure}[htbp]
\begin{center}
\includegraphics[width=7cm]{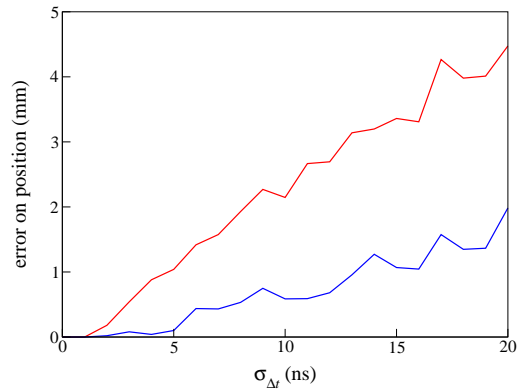}
\caption{Error on the hit location as a function of the standard deviation of the time shift jitter. The signal corresponds to a point located at the center of a segment and is affected by a 1\% noise. The red curve shows the average results using, as minimization criterion, the residue and the blue line the average results using the chi-square, Eqs. (\ref{Eq chi2}),(\ref{Eq sigma2s}).}
\label{Fig ecd_b1_340}
\end{center}
\end{figure}

To compare the results given by the two comparison criteria, we have simulated signals affected by a 1\% noise and by time shifts sorted at random in a Gaussian distribution with standard deviation $\sigma_{\Delta t}$. The location of the hit is then determined with a grid search algorithm using both minimization criteria. The qualities of the results are characterized by the error on the location of the hit. A first test is performed considering a given point at the center of a segment, i.e. belonging to a region of the crystal where the sensitivity (of the signal shape to the location of the hit) is low. This means that the results obtained for this particular test signal are among the worst that can be expected from the crystal. The results are shown in Fig. \ref{Fig ecd_b1_340} which represent the error averaged over noise using the residue (red curve) and the chi-square (blue curve) criteria. As can be seen, the results for the chi-square are much better. Up to a time shift jitter of 5 ns no error is made on the location of the hit. For higher time shifts, the error is more than twice as large for the residue criterion.

\begin{figure}[htbp]
\begin{center}
\includegraphics[width=7cm]{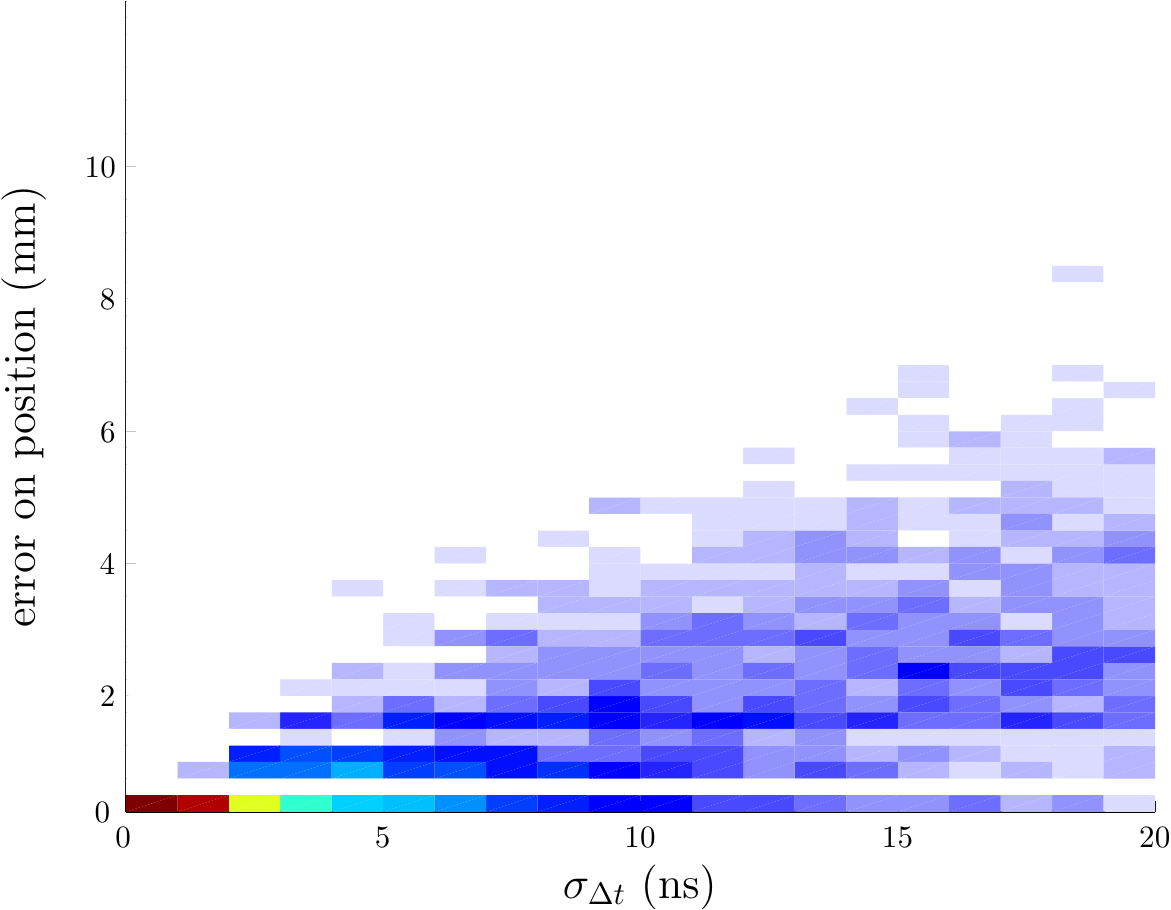}
\includegraphics[width=7cm]{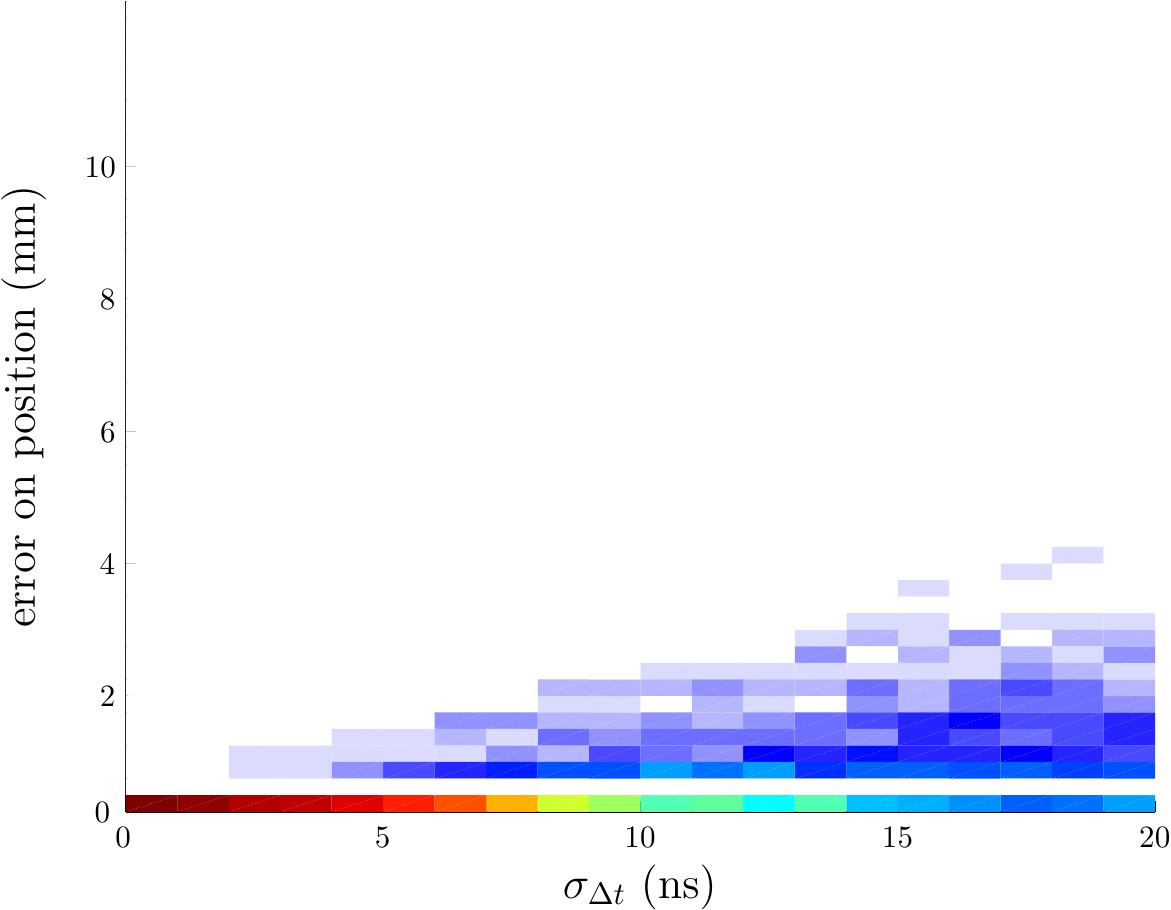}
\caption{Error on the hit location as a function of the standard deviation of the time shift jitter for the whole set of signals with 1\% noise. Upper plot: using the residue criterion. Lower plot: using the chi-square criterion.}
\label{Fig ecd_b1_o1}
\end{center}
\end{figure}

The results obtained for the whole set of signals are shown in Fig. \ref{Fig ecd_b1_o1} leading to the same conclusions.

In spite of its simplicity, this method gives a very good improvement to the location of gamma-ray interactions. Moreover, the estimate of the signal standard deviation given in Eq. (\ref{Eq sigma2s}), depends only on the reference basis and on the standard deviations of the noise and the time shift. Thus, the standard deviation distributions for each of the basis signals can be pre-calculated off-line. The increase in computer time for the on-line calculation of Eq. (\ref{Eq chi2}) instead of Eq. (\ref{Eq R2}) is a low price to pay.

\section{Algebraic time shift correction}

In the following, we search for the time shift $\Delta t$ to impose to a reference signal $\bfsz$ so that it matches the signal $\bfs$ delivered by the detector. In other words, we search for $\Delta t$ such that:

\begin{equation}
\label{Eq R2}
R^2 = \sum_{t} \left(s_{0}(t+\Delta t) - s(t) \right)^2\,,
\end{equation}

\noindent is minimum. Here, we use the residue instead of the chi-square since, unlike in the previous Section, we need a criterion sensitive to the time jitter. Assuming that the time shift is small enough, the shifted signal can be approximated by:

\begin{equation}
\label{Eq o2}
s_{0}(t+\Delta t) = s_{0}(t) + \frac{\partial s_{0}}{\partial t}\ \Delta t+ \frac{1}{2}\frac{\partial^2 s_{0}}{\partial t^2}\ \Delta t^2+ \frac{1}{6}\frac{\partial^3 s_{0}}{\partial t^3}\ \Delta t^3...
\end{equation}

The first three derivatives of the signal can be estimated by:
\begin{equation}
\begin{array}{l}
p(t)=\frac{\partial s_{0}}{\partial t} =  \frac{s_{0}(t+1)-s_{0}(t-1)}{2\,\ts}\\
q(t) = \frac{\partial^{2} s_{0}}{\partial t^2}= \frac{s_{0}(t+2)-2s_{0}(t)+s_{0}(t-2)}{4\,\ts^2}\\
r(t)= \frac{\partial^{3} s_{0}}{\partial t^3} =
  \frac{s_{0}(t+3)-3s_{0}(t+1)+3s_{0}(t-1)-s_{0}(t-3)}{8\,\ts^3}
\end{array}
\end{equation}

The developing of the residue to the second order gives:

\begin{equation}
\begin{array}{lll}
\label{Eq R2 o2}
R^2(\Delta t) =& & (\bfsz-\bfs)^2\\
& + &2\, \bfp . (\bfsz-\bfs)\, \Delta t\\
&+ &[\bfp^2+\bfq . (\bfsz-\bfs)]\, \Delta t^2\,.
\end{array}
\end{equation}

The minimum of this function\footnote{As we need $R^2(\Delta t)$ to show a minimum, it must be developed at least to the second order.} is reached for:

\begin{equation}
\label{Eq Dt o2}
\Delta t =-\frac{c-\bfp . \bfs}{b-\bfq . \bfs}\,,
\end{equation}

\noindent where $c$ and $b$ are constants that depend only on $\bfsz$:

\begin{eqnarray}
c &=& \bfp . \bfsz\,,\\
b &=& \bfp^2 + \bfq . \bfsz\,.
\end{eqnarray}

Thus, they can be pre-calculated.

In order to correct longer time shifts, the third order term of the residue can also be considered. This term, to be added to Eq. (\ref{Eq R2 o2}), reads:

\begin{equation}
\begin{array}{ll}
\label{Eq R2 o3}
+[\frac{1}{3}\,\bfr . (\bfsz-\bfs)+\bfp . \bfq]\,\Delta t^3\,.
\end{array}
\end{equation}

Once more, setting the derivative to zero gives the value of the time shift that minimizes the residue:

\begin{equation}
\label{Eq Dt o3}
\Delta t =\frac{-(b-\bfq . \bfs)+\sqrt{(b-\bfq . \bfs)^2-2(a-\bfr . \bfs)(c-\bfp . \bfs)}}{a-\bfr . \bfs}
\end{equation}

\noindent where $a$ is also a constants that depends only on $\bfsz$:

\begin{eqnarray}
a &=& 3\, \bfp . \bfq + \bfr . \bfsz\,.
\end{eqnarray}

The quality of the match of two signals can also be measured using their scalar product \cite{Dox,Des1}. The resulting time shift is $\Delta t = -(\bfp . \bfs)/(\bfq . \bfs)$. The tests show that the agreement with the true time shift is much better using the minimum of the residue than using the maximum of the scalar product. Thus only the results obtained with the residue are shown in the next Section.

\subsection{Results}

\subsubsection{Validation criteria}

In order to validate this time shift correction method, we consider two criteria. The first one is obviously the error on the time shift defined as the difference between the real time shift and the values found by Eqs. (\ref{Eq Dt o2}) and (\ref{Eq Dt o3}). The second criterion measures the error on the location of the hit. In order to evaluate the improvements, we will compare the errors, with and without time shift correction, using grid search.

\subsubsection{Error on the time shift evaluation}

We first consider as test signal a meta-signal corresponding to the center of a segment. This test signal was translated by 1~ns time steps from 1 to 100~ns. Three deposited energies are considered: 300 keV (the noise is 1\% of the maximum amplitude of the segment signal), 100 keV (3\% noise) and 30 keV (10\% noise).

\begin{figure}[htbp]
\begin{center}
\includegraphics[width=5.5cm]{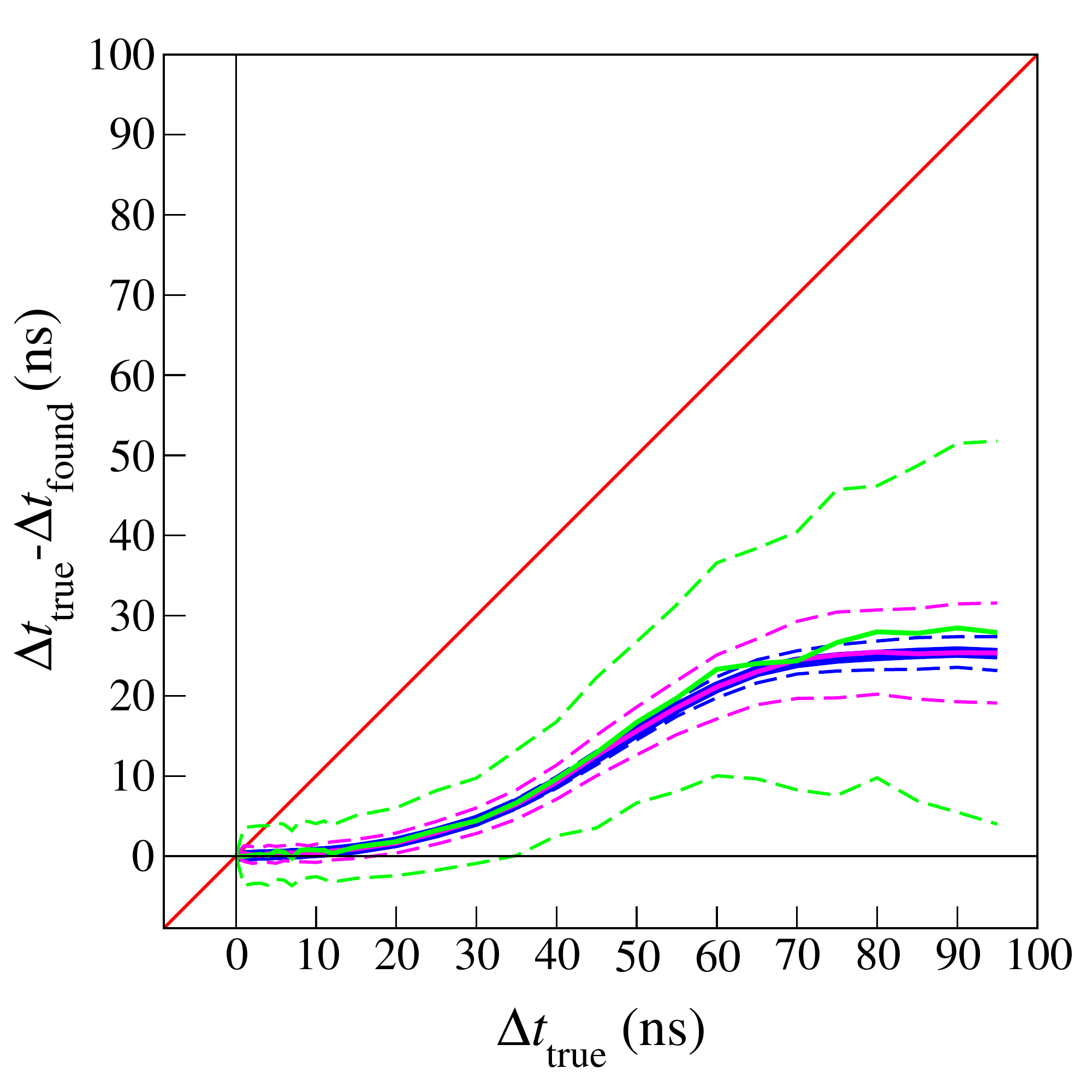}
\includegraphics[width=5.5cm]{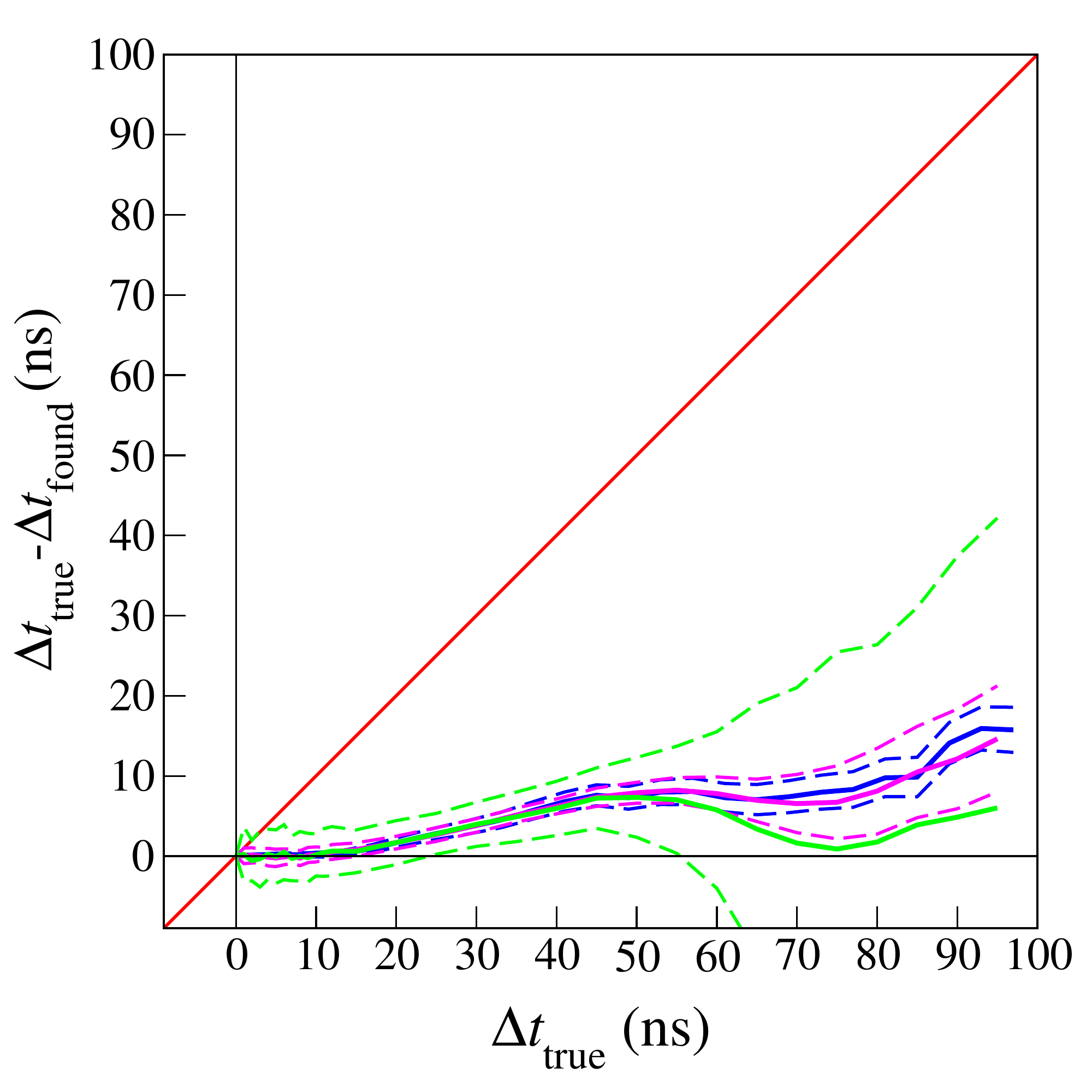}
\caption{Error on the time shift correction as a function of the imposed time shift for a point located at the center of a segment. Upper plot: second order correction, Eq. (\ref{Eq Dt o2}). Lower plot: third order correction, Eq. (\ref{Eq Dt o3}). The red line shows the result when no correction is applied. The bold lines correspond to the values averaged over noise and the dashed lines to one standard deviation from the mean (blue: 1\% noise, violet: 3 \% noise, green: 10\% noise).}
\label{Fig et_340}
\end{center}
\end{figure}

The results obtained by the second order correction, Eq. (\ref{Eq Dt o2}), for this signal are presented in the upper plot of Fig. \ref{Fig et_340}. The time shift error $\Dtt-\Dtf$ is shown as a function of the imposed time shift for different amount of noise. The bold lines represent the mean time shift error averaged over noise and the upper and lower dashed lines the standard deviations. As can be seen, the results are very good. In this case, the time shift correction is better than 1~ns up to time shifts equal to 15~ns and the error is less than 10\% up to 25~ns. Remarkably, the average correction is not sensitive to the amount of noise, even for a large value.

The results for the third order correction, Eq. (\ref{Eq Dt o3}), are show in the lower plot of Fig. \ref{Fig et_340}. In the left part of the curve, the quality of the correction is the same as for second order correction. As expected, the time shift correction is better for large values of $\Delta t$. 

In these plots, we have presented the results obtained in a large range of time shifts. In fact, when the shift is longer than some sample times, other efficient correction procedures may be used. The residue minimization method is particularly suited for small, intra-sample, adjustments. Thus, in the following, we will focus on the left part of the figures.

\begin{figure}[htbp]
\begin{center}
\includegraphics[width=7cm]{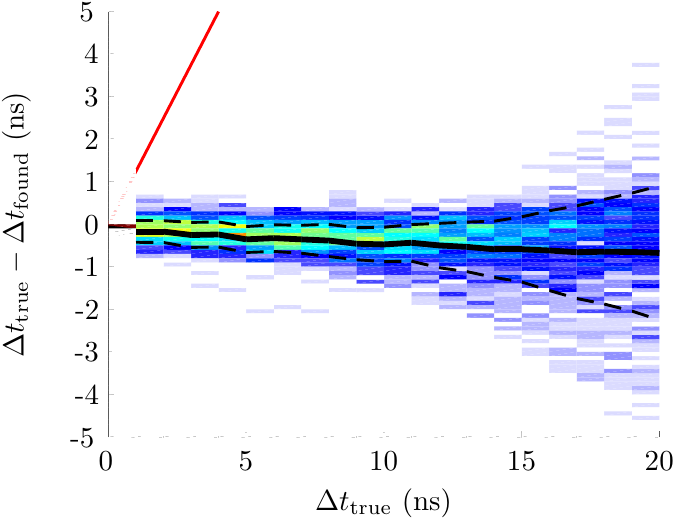}
\includegraphics[width=7cm]{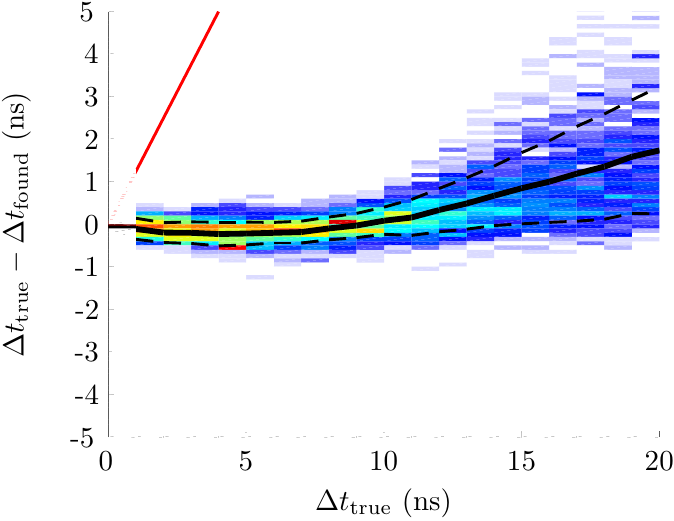}
\caption{Error on the time shift correction $\Dtt-\Dtf$ as a function of the imposed time shift for the whole set of signals. The black full line corresponds to the average value obtained for the test signals and the dashed lines to one standard deviation from the mean. The red line shows the result without correction. Upper plot: second order correction. Lower plot: third order correction.}
\label{Fig et_ac_b1_o1}
\end{center}
\end{figure}

We now consider the whole set of basis signals. Each of them is successively used as test signal being added a 1\% noise and shifted by 1~ns steps. The results of the correction are shown in Fig. \ref{Fig et_ac_b1_o1}. The upper plot corresponds to the second order correction. In the first sample, that is for time shifts lower than 10~ns, the error on the correction is lower than 0.25~ns in 68\% of the cases. For a two-sample shift, the standard error of the correction is better than 10\%. Inside the first sample, the correction obtained with the third order method is even better than with the second order one. A small positive bias is visible in the second sample.

The final goal of pulse shape analysis, in the case of segmented HPGe detector, is to determine the location of the hits. As for the chi-square method, we now test the improvement on the position.

\subsubsection{Error on the hit location}

\begin{figure}[htbp]
\begin{center}
\includegraphics[width=7cm]{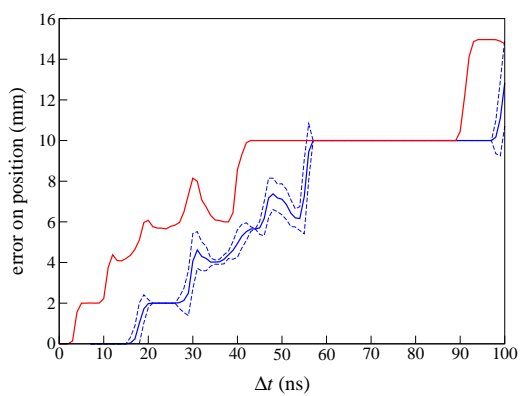}
\caption{Error on the location of a hit as a function of the imposed time shift. The signal is the same as in Fig. \ref{Fig ecd_b1_340} and  \ref{Fig et_340}. The red curve shows the average results without correction and the blue curve shows the average results after second order correction (the dotted lines correspond to the mean plus or minus the standard deviation).}
\label{Fig ed_o1_340}
\end{center}
\end{figure}

The correction of the time shift allows a better location of the gamma-ray interactions. In the case of Fig. \ref{Fig ed_o1_340}, when no correction is applied, a 4~ns shift may result in a 2~mm error on the position. On the other hand, a second order correction allows to have a perfect hit location up to a time shift of 15~ns, (1 \% noise) which is largely sufficient for signal decomposition purpose.

\begin{figure}[htbp]
\begin{center}
\includegraphics{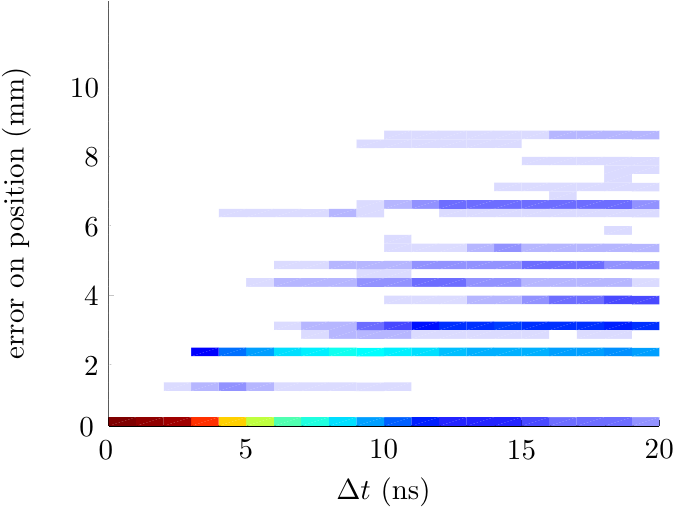}
\includegraphics{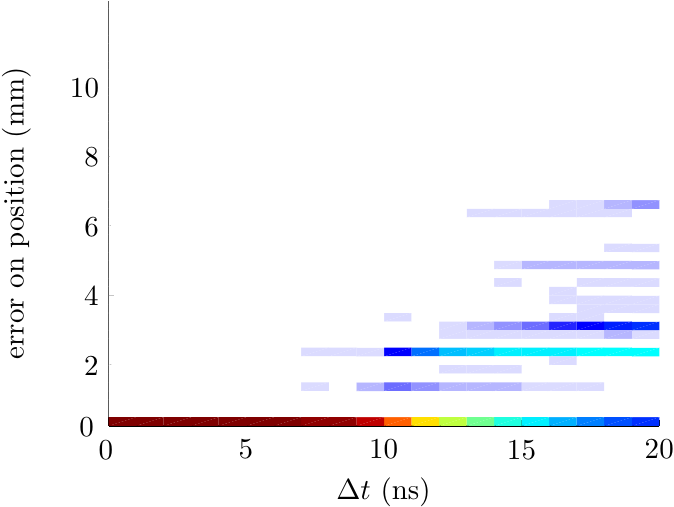}
\caption{Error on the hit location as a function of the imposed time shift for the whole set of signals with 1\% noise. Upper plot: no time shift correction. Lower plot: second order time shift correction.}
\label{Fig ed_b1_o1}
\end{center}
\end{figure}

The results obtained for the whole set of test signals are shown in Fig. \ref{Fig ed_b1_o1}. The upper plot corresponds to the error on the location of the hit when no time shift correction is applied and the lower plot present the result after second order time correction. When the correction is applied, the grid search makes almost no location error when the time shifts are lower than the sample time.

\section{Conclusions}

This paper proposes two different methods for the correction of signal time shifts. The robustness of these methods is connected to the fact that, unlike CFD threshold techniques for example, they use the whole signal to evaluate the shift, thus the amount of relevant information is high and the negative influence of noise is minimized.

The first method allows the comparison of basis signals with detected signals affected by white noise and random time shift. It improves the confidence on the determination of the best-match basis signal. In our example, this induces a much better location of the gamma-ray interactions in a germanium detector. The increase of computing time, due to the use of a chi-square rather than a residue, is largely smaller than what would have been necessary to make a minimization by systematically comparing the detected signal with a set of time shifted versions of the reference signals.

In the second part of this paper, we have proposed an algebraic evaluation of the time shift based on the Taylor polynomial of the reference signal and on the minimization of the residue. The reliability of this method is comparable with other methods when the shift is greater than a few samples. For intra-sample time shift, the results are excellent (the standard deviation is lower than one fortieth of the sample time for a 1\% noise). The choice between second and third order corrections depends on the application and on the available computing time in case of on-line correction. The second order correction entails mostly the calculation of two scalar products, Eq. (\ref{Eq Dt o2}), and the third order correction, the calculation of three scalar products and a square root, Eq. (\ref{Eq Dt o3}). In principle, third order correction gives better time shift estimates, but, as it entails the calculation of higher degrees functions, the uncertainty on the result is larger. However, whatever the order of the correction, the method is much faster than the traditional method of exhaustive time-shift trials.

\end{document}